\documentclass[12pt,preprint]{aastex}

\received{}
\accepted{}

\slugcomment{The Astrophysical Journal}

\shorttitle{WHY IS ZEL'DOVICH APPROXIMATION SO ACCURATE?}
\shortauthors{YOSHISATO ET AL.}

\begin{document}

\title{WHY IS THE ZEL'DOVICH APPROXIMATION SO ACCURATE?}

\author{AYAKO YOSHISATO \altaffilmark{1} and MASAHIRO MORIKAWA}
\affil{Department of Physics, Ochanomizu University, Otsuka 2-2-1, Bunkyo, Tokyo 112-8610, Japan; ayako@cosmos.phys.ocha.ac.jp}

\author{NAOTERU GOUDA}
\affil{National Astronomical Observatory of Japan, Osawa 2-21-1, Mitaka, Tokyo 181-8588, Japan}

\author{HIDEAKI MOURI}
\affil{Meteorological Research Institute, Nagamine 1-1, Tsukuba 305-0052, Japan}

\altaffiltext{1}{Research Institute of Systems Planning, Sakuragaoka-cho 2-9, Shibuya, Tokyo 150-0031, Japan}

\begin{abstract}

Why does the Zel'dovich approximation (ZA) work well to describe gravitational collapse in the universe? This problem is examined by focusing on its dependence on the dimensionality of the collapse. The ZA is known to be exact for a one-dimensional collapse. We show that the ZA becomes progressively more accurate in the order of three-, two-, and one-dimensional collapses. Furthermore, using models for spheroidal collapse, we show that the ZA remains accurate in all collapses, which become progressively lower dimensional with the passage of time. That is, the ZA is accurate because the essence of the gravitational collapse is incorporated in the ZA.\end{abstract}

\keywords{cosmology: theory    --- 
          large scale structure of universe}

\section{INTRODUCTION}
\label{s1}

For gravitational collapse in the universe, Zel'dovich (1970) proposed an approximation, the Zel'dovich approximation (ZA). The ZA is a Lagrangian approximation based on the motion of a fluid element. If its initial position is $\mbox{\boldmath $q$}$ in comoving coordinates, the Eulerian position $\mbox{\boldmath $x$}(\mbox{\boldmath $q$},t)$ at time $t$ is
\begin{equation}
\label{eq1}
\mbox{\boldmath $x$}(\mbox{\boldmath $q$},t) 
=
\mbox{\boldmath $q$} + 
{\bf \Psi}(\mbox{\boldmath $q$},t).
\end{equation}
The displacement ${\bf \Psi}(\mbox{\boldmath $q$},t)$ satisfies the Lagrangian equation of motion (Peacock 1999, \S15.2):
\begin{equation}
\label{eq2}
\frac{\partial^2 {\bf \Psi}}{\partial t^2} + 2H\frac{\partial {\bf \Psi}}{\partial t} 
=
- {\bf \nabla}_x \Phi.
\end{equation}
Here $H(t)$ is the Hubble parameter. The gravitational potential $\Phi(\mbox{\boldmath $x$},t)$ is obtained from the density contrast $\delta(\mbox{\boldmath $x$}, t) = \det (\partial x_i / \partial q _j )^{-1} -1$ and the density parameter $\Omega (t)$ via the Poisson equation
\begin{equation}
\label{eq3}
\nabla_{x}^{\,2} \Phi = \frac{3}{2} H^2 \Omega \delta .
\end{equation}
From the first-order perturbation to equations (\ref{eq2}) and (\ref{eq3}), Zel'dovich (1970; see also  Shandarin \& Zel'dovich 1989; Sahni \& Coles 1995) assumed that the displacement ${\bf \Psi}$ scales with the linear growth factor $D(t)$:  
\begin{equation}
\label{eq4}
{\bf \Psi} = - \frac{2 D}{3 H^2_{\rm in} \Omega _{\rm in}} {\bf \nabla} _q \Phi _{\rm in}.
\end{equation}
Here the subscript ``in'' indicates a value at the initial time $t_{\rm in}$. The linear growth factor $D$ satisfies the evolution equation
\begin{equation}
\label{eq5}
\frac{d^2 D}{dt^2} 
+
2 H \frac{dD}{dt} 
=
\frac{3}{2} H^2 \Omega D.
\end{equation}
Hence, the displacement ${\bf \Psi}$ is assumed to satisfy the following equation:
\begin{equation}
\label{eq6}
\frac{\partial ^2 {\bf \Psi}}{\partial t^2} 
+
2 H \frac{\partial {\bf \Psi}}{\partial t} 
=
\frac{3}{2} H^2 \Omega {\bf \Psi}.
\end{equation}
Equation (\ref{eq4}) or (\ref{eq6}) provides an approximation for the gravitational collapse, until the shell crossing where fluid elements from different initial positions reach the same position in Eulerian coordinates. The approximation applies to any cosmological model.

The ZA is used for various topics in observational cosmology, e.g., the construction of the initial conditions in an $N$-body simulation (Doroshkevich et al. 1980; Bertschinger 1998), the estimation of the initial conditions of the universe from observations of galaxies in the present universe (Dekel 1994), and the analytical study of the pairwise peculiar velocities of galaxies (Yoshisato et al. 2003). The ZA is also used to develop the adhesion approximation (Gurbatov et al. 1989), the truncated ZA (Coles et al. 1993), and the post-, post-post-, and Pad\'e-ZAs (Bernardeau 1994; Buchert 1994; Munshi et al. 1994; Bouchet et al. 1995; Catelan 1995; Matsubara et al. 1998). Thus, the ZA plays an important role.

Despite its simplicity, the ZA is accurate even in the quasi-nonlinear regime, provided that the shell crossing is unimportant as at early times or at large scales.\footnote{
Yoshisato et al. (2003) proposed a method for determining the scale range where the shell crossing is unimportant at a given time.} 
This accuracy has been demonstrated in statistical comparison with an $N$-body simulation (Coles et al. 1993; Kofman et al. 1994) and with a high-order Eulerian perturbation theory (Bernardeau et al. 1994; Munshi \& Starobinsky 1994).

Why is the ZA so accurate? The usual explanation invokes the fact that the Lagrangian description is intrinsically nonlinear in the density field (Catelan 1995; Bernardeau et al. 2002). The density contrast $\delta$ is obtained from the displacement ${\bf \Psi}$. Even if the displacement is small and lies in the linear regime, the corresponding density contrast could be large and lie in the nonlinear regime.

However, the ZA's accuracy depends on the shape of the overdense region. We believe that this dependence is the most essential (Yoshisato et al. 1998; Peacock 1999, \S15.8). The past comparison of the ZA with an exact solution was made only for a spherical collapse (Munshi et al. 1994; Sahni \& Shandarin 1996). Yoshisato et al. (1998) made the comparison for spheroidal collapses, which are more usual situations in the real universe. It was found that, for a given value of the density contrast, the ZA is more accurate in a lower dimensional collapse.\footnote{
Throughout this paper, as in Yoshisato et al. (1998), we use the dimension to characterize the shape of the collapse, which is assumed to occur in a three-dimensional space. The one-, two-, and three-dimensional collapses respectively correspond to planar, cylindrical, and spherical collapses.}

Zel'dovich (1970) noticed that, in the ZA equation (\ref{eq6}), the acceleration term $- {\bf \nabla}_x \Phi$ in the true equation of motion (\ref{eq2}) has been replaced with the linear term for the displacement $3 H^2 \Omega {\bf \Psi}/2$. This gravitational acceleration of the ZA is identical to the true gravitational acceleration in a one-dimensional collapse (Buchert 1989), for which the ZA is exact (Doroshkevich et al. 1973).

Here we analytically show that the ZA's gravitational acceleration approaches the true gravitational acceleration in the order of three-, two-, then one-dimensional collapses (\S\ref{s2}). We also numerically show that any spheroidal collapse becomes progressively lower dimensional with the passage of time. The ZA's gravitational acceleration remains close to the true gravitational acceleration (\S\ref{s3}). Then we discuss why the ZA is accurate (\S\ref{s4}).

\section{DIMENSIONAL DEPENDENCE OF EQUATION OF MOTION}
\label{s2}
The Lagrangian equations of motion (\ref{eq2}) are derived for one-dimensional planar, two-dimensional cylindrical, and three-dimensional spherical collapses. We compare these equations with the ZA equation (\ref{eq6}).

\subsection{One-dimensional Planar Collapse}
\label{s21}

When the collapse is planar, the gravitational potential $\Phi (x,t)$ is obtained using Green's function $3 H^2 \Omega \vert x \vert /4$ for the one-dimensional Poisson equation (\ref{eq3}):
\begin{equation}
\label{eq7}
\Phi
= \frac{3}{4} H^2 \Omega \int^{\infty}_{-\infty} \delta (x',t) \vert x - x' \vert dx'.
\end{equation}
The gravitational acceleration of the fluid element at position $x$ is 
\begin{equation}
\label{eq8}
-\frac{\partial \Phi}{\partial x}
=
-\frac{3}{4} H^2 \Omega
\left[ \int^x_{-\infty} \delta (x',t) dx' - \int^{\infty}_x \delta (x',t) dx' \right].
\end{equation}
This gravitational acceleration is proportional to the difference of the integrated density contrast between both sides of the position $x$. The reason is that, in a one-dimensional collapse, the gravity is independent of the distance. We rewrite equation (\ref{eq8}) with the displacement of the fluid element $\Psi = x - q$. Here $q$ is the initial position. Until the shell crossing, mass conservation leads to
\begin{eqnarray}
\label{eq9}
\int^x_{-\infty} 1+\delta (x',t) dx' & = & \int ^q _{-\infty} dx', \nonumber \\
\int^{\infty}_x  1+\delta (x',t) dx' & = & \int ^{\infty}_q   dx'.
\end{eqnarray}
Here we set the initial density contrast to zero. It follows from equation (\ref{eq9}) that the integrated density contrast is proportional to the displacement $\Psi$:
\begin{eqnarray}
\label{eq10}
\int^x_{-\infty} \delta (x',t) dx' & = & \int^q_{-\infty} dx' - \int^x_{-\infty} dx'
                                     = q-x = - \Psi ,
\nonumber \\
\int^{\infty}_x  \delta (x',t) dx' & = & \int^{\infty}_q  dx' - \int^{\infty}_x  dx' 
                                     = x-q = \Psi.
\end{eqnarray}
The gravitational acceleration is
\begin{equation}
\label{eq11}
- \frac{\partial \Phi}{\partial x} = \frac{3}{2} H^2 \Omega \Psi.
\end{equation}
Thus, in the one-dimensional planar collapse, the true gravitational acceleration is identical to the ZA's gravitational acceleration.

\subsection{Two-dimensional Cylindrical Collapse}
\label{s22}

When the collapse is cylindrical, we use Green's function $3 H^2 \Omega \ln \vert \mbox{\boldmath $x$} \vert / 4 \pi$ for the two-dimensional Poisson equation (\ref{eq3}). The gravitational acceleration of the fluid element at position $\mbox{\boldmath $x$}$ is
\begin{equation}
\label{eq12}
-\frac{\partial \Phi}{\partial \vert \mbox{\boldmath $x$} \vert} 
=
-\frac{3}{2} H^2 \Omega \frac{1}{\vert \mbox{\boldmath $x$} \vert} 
\int^{\vert \mbox{\boldmath $x$} \vert}_0 \delta (r',t) r' dr' .
\end{equation}
This acceleration is proportional to $\vert \mbox{\boldmath $x$} \vert ^{-1}$ times the density contrast integrated over the radius $\vert \mbox{\boldmath $x$} \vert$, in accordance with Gauss's theorem (Binney \& Tremaine 1987, \S2.0). The gravity in a two-dimensional collapse is proportional to the inverse of the distance. We rewrite equation (\ref{eq12}) with the initial position of the fluid element $\mbox{\boldmath $q$}$. Until the shell crossing, mass conservation leads to
\begin{equation}
\label{eq13}
\int^{\vert \mbox{\boldmath $x$} \vert}_0 [1+\delta ( r',t )] r'dr'
=
\int^{\vert \mbox{\boldmath $q$} \vert}_0 r' dr' .
\end{equation}
Then, as in \S\ref{s21}, the gravitational acceleration is 
\begin{equation}
\label{eq14}
-\frac{\partial \Phi}{\partial \vert \mbox{\boldmath $x$} \vert} 
=
\frac{3}{4} H^2 \Omega
\frac{ \vert \mbox{\boldmath $x$} \vert ^2 - 
\vert \mbox{\boldmath $q$} \vert ^2 }{\vert \mbox{\boldmath $x$} \vert}.
\end{equation}
Since the displacement $\Psi$ is $\vert \mbox{\boldmath $x$} \vert-\vert \mbox{\boldmath $q$} \vert$, this true gravitational acceleration differs from the ZA's gravitational acceleration $3 H^2 \Omega \Psi /2$ by a factor $(\vert \mbox{\boldmath $x$} \vert+\vert \mbox{\boldmath $q$} \vert)/2 \vert \mbox{\boldmath $x$} \vert$.

\subsection{Three-dimensional Spherical Collapse}
\label{s23}

When the collapse is spherical, we use Green's function $- 3 H^2 \Omega / 8 \pi r$ for the three-dimensional Poisson equation (\ref{eq3}). The gravitational acceleration of the fluid element at position $\mbox{\boldmath $x$}$ is
\begin{equation}
\label{eq15}
-\frac{\partial \Phi}{\partial \vert \mbox{\boldmath $x$} \vert}
= 
-\frac{3}{2} H^2 \Omega \frac{1}{\vert \mbox{\boldmath $x$} \vert ^2}
\int^{\vert \mbox{\boldmath $x$} \vert}_0 \delta (r',t) r'^2 dr'.
\end{equation}
This acceleration is proportional to $\vert \mbox{\boldmath $x$} \vert ^{-2}$ times the density contrast integrated over the radius $\vert \mbox{\boldmath $x$} \vert$, in accordance with Gauss's theorem (Binney \& Tremaine 1987, \S2.0). The gravity in a three-dimensional collapse is proportional to the inverse of the distance squared. We rewrite equation (\ref{eq15}) with the initial position of the fluid element $\mbox{\boldmath $q$}$. Until the shell crossing, mass conservation leads to
\begin{equation}
\label{eq16}
\int^{\vert \mbox{\boldmath $x$} \vert}_0 [1+ \delta (r',t)] r'^2 dr'
=
\int^{\vert \mbox{\boldmath $q$} \vert}_0  r'^2 dr' .
\end{equation}
Then, as in \S\ref{s21} and \ref{s22}, the gravitational acceleration is 
\begin{equation}
\label{eq17}
-\frac{\partial \Phi}{\partial \vert \mbox{\boldmath $x$} \vert}
=
\frac{1}{2} H^2 \Omega
\frac{\vert \mbox{\boldmath $x$} \vert ^3 - \vert \mbox{\boldmath $q$} \vert ^3}
     {\vert \mbox{\boldmath $x$} \vert ^2} .
\end{equation}
Since the displacement $\Psi$ is $\vert \mbox{\boldmath $x$} \vert-\vert \mbox{\boldmath $q$} \vert$, this true gravitational acceleration differs from the ZA's gravitational acceleration $3 H^2 \Omega \Psi /2$ by a factor $(\vert \mbox{\boldmath $x$} \vert^2 + \vert \mbox{\boldmath $x$} \vert \vert \mbox{\boldmath $q$} \vert + \vert \mbox{\boldmath $q$} \vert^2)/3 \vert \mbox{\boldmath $x$} \vert^2$. 

\subsection{Dimensional Dependence}
\label{s24} 

Summarizing the equations of motion for one-dimensional planar, two-dimensional cylindrical, and three-dimensional spherical collapses (eqs. [\ref{eq11}], [\ref{eq14}], and [\ref{eq17}]), we have 
\begin{equation}
\label{eq18}
\frac{\partial ^2 \Psi}{\partial t^2 }+2H \frac{\partial \Psi}{\partial t}
=
\frac{3}{2} H^2 \Omega
\frac{1}{\vert \mbox{\boldmath $x$} \vert ^{n-1}}
\frac{\vert \mbox{\boldmath $x$} \vert ^n - \vert \mbox{\boldmath $q$} \vert ^n}{n}
,
\end{equation}
with the displacement $\Psi = \vert \mbox{\boldmath $x$} \vert-\vert \mbox{\boldmath $q$} \vert$ and the dimension $n = 1$, 2, or 3. The factor $1/\vert \mbox{\boldmath $x$} \vert ^{n-1}$ reflects the dependence of the gravity on the distance $\vert \mbox{\boldmath $x$} \vert$. The factor $(\vert \mbox{\boldmath $x$} \vert ^n - \vert \mbox{\boldmath $q$} \vert ^n)/n$ reflects the change of the integrated density contrast in the course of the gravitational collapse.

We compare equation (\ref{eq18}) with the ZA equation (\ref{eq6}), in order to study the ZA's accuracy. When the displacement is small, these two equations are almost identical. The ZA is accurate. Even when the displacement is large, in a lower dimensional collapse, the gravity $1/\vert \mbox{\boldmath $x$} \vert ^{n-1}$ is more independent of the distance $\vert \mbox{\boldmath $x$} \vert$, and the change of the integrated density contrast $(\vert \mbox{\boldmath $x$} \vert ^n - \vert \mbox{\boldmath $q$} \vert ^n)/n$ is closer to the displacement $\vert \mbox{\boldmath $x$} \vert - \vert \mbox{\boldmath $q$} \vert$. The ZA becomes progressively more accurate in the order of three-, two-, and one-dimensional collapses. For the one-dimensional collapse, the ZA equation (\ref{eq6}) is exact. These results are independent of the functional forms of $H(t)$ and $\Omega (t)$ and thus apply to any cosmological model.

\section{SPHEROIDAL COLLAPSE}
\label{s3}

During a real gravitational collapse, the overdense region changes its shape. The effect of this change on the ZA's accuracy, especially in the gravitational acceleration, is studied by using the gravitational collapse of a homogeneous spheroid as an idealized model (see also Yoshisato et al. 1998).

When pressure is negligible, the gravitational collapse increases the ellipticity of the spheroid (Lin et al. 1965). While an initially oblate spheroid forms a plane-like structure, an initially prolate spheroid forms a line-like structure. This is because the gradient of the gravitational potential generated by the spheroid is larger along the minor axis of the spheroid than along the major axis. The gravitational acceleration is stronger and the gravitational collapse is more significant along the minor axis (Binney 1977). Since the collapse becomes progressively lower dimensional with the passage of time, the ZA is expected to remain accurate.

\subsection{Equation of Motion}
\label{s31}

The density contrast in the presence of a homogeneous ellipsoid is determined by the half-lengths of its principal axes $\alpha_i(t)$:
\begin{equation}
  \label{eq19}
  \delta(\mbox{\boldmath $x$}, t) =
  \delta_{\rm e}(t)
  \Theta\left[
    1 - \frac{x_1^2}{\alpha_1^2(t)}
      - \frac{x_2^2}{\alpha_2^2(t)}
      - \frac{x_3^2}{\alpha_3^2(t)}
  \right].
\end{equation}
Here $\Theta (x)$ is a step function: $\Theta (x) = 1$ for $x \ge 0$ and 0 for $x < 0$. The density contrast within the ellipsoid $\delta_{\rm e}(t)$ is 
\begin{equation}
  \label{eq20}
  \delta_{\rm e}(t) =
  (1 + \delta_{\rm in})
  \frac{\alpha_{1,{\rm in}}\alpha_{2,{\rm in}}\alpha_{3,{\rm in}}}
    {\alpha_1(t) \alpha_2(t) \alpha_3(t)}
  - 1.
\end{equation}
We adopt spheroidal symmetry, $\alpha_1(t)=\alpha_2(t)$. Then the Poisson equation (\ref{eq3}) yields the gravitational potential $\Phi  (\mbox{\boldmath $x$}, t)$ as
\begin{equation}
  \label{eq21}
  \Phi
= \frac{3}{8} H^2 \Omega \delta_{\rm e}
  \sum_{i=1}^3 A_i x_i^{\,2},
\end{equation}
with
\begin{equation}
   \label{eq22}
   A_1(t) = A_2(t) = \frac{2}{3} [1 + h(t)], \quad
   A_3(t) = \frac{2}{3} [1 - 2 h(t)].
\end{equation}
For an oblate case $\left(\alpha_1 = \alpha_2 > \alpha_3\right)$, we have
\begin{equation}
\label{eq23}
  h(t) =
       \frac{3}{2} \frac{\sqrt{1 - e^2}}{e^3} \sin^{-1} e -
       \frac{3 - e^2}{2 e^2} ,
\end{equation}
with the ellipticity
\begin{equation}
\label{eq24}
  e(t) = \left[ 1 - \frac{\alpha_3(t)^2}{\alpha_1(t)^2} \right] ^{1/2}.
\end{equation}
For a prolate case $\left(\alpha_1 = \alpha_2 < \alpha_3\right)$, we have\footnote{The eqs. (33) and (36) in Yoshisato et al. (1998) that correspond to our equations (\ref{eq20}) and (\ref{eq25}) have typographical errors.}
\begin{equation}
\label{eq25}
  h(t) =
      \frac{3}{4} \frac{1 - \bar{e}^2}{\bar{e}^3}
      \ln\left(\frac{1 - \bar{e}}{1 + \bar{e}}\right) +
      \frac{3 - 2 \bar{e}^2}{2 \bar{e}^2} ,
\end{equation}
with the ellipticity
\begin{equation}
\label{eq26}
      \bar{e}(t) = \left[ 1 - \frac{\alpha_1(t)^2}{\alpha_3(t)^2} \right] ^{1/2}.
\end{equation}
These equations are derived in Binney \& Tremaine (1987, \S2.3). We consequently obtain the equation of motion (\ref{eq2}) for the half-lengths $\alpha _i(t)$
\begin{equation}
  \label{eq27}
  \frac{d^2 \alpha _i}{dt^2} + 2H \frac{d \alpha _i}{dt} = 
  - \frac{3}{4} H^2 \Omega \delta_{\rm e} A_i \alpha _i.
\end{equation}
The evolution of the half-lengths $\alpha _i(t)$ describes the evolution of the homogeneous spheroid.

\subsection{ZA}
\label{s32}

Equations (\ref{eq4}) and (\ref{eq21}) yield the equation for the evolution of the half-lengths $\alpha _i(t)$ in the ZA:
\begin{equation}
\label{eq28}
\alpha _i - \alpha _{i,{\rm in}} 
=
- \frac {1}{2} D \delta _{\rm e, in} A_{i,{\rm in}} \alpha _{i,{\rm in}} .
\end{equation}
The ZA equation of motion (\ref{eq6}) is
\begin{equation}
\label{eq29}
\frac{d^2 \alpha _i}{dt^2} + 2H \frac{d \alpha _i}{dt} 
= \frac{3}{2} H^2 \Omega (\alpha _i - \alpha _{i,{\rm in}}) .
\end{equation}
The factor $A_{i,{\rm in}}$ is larger along the minor axis of the spheroid than that along the major axis. Hence, also in the ZA, the gravitational collapse is more significant along the minor axis. The ellipticity $e(t)$ or $\bar{e}(t)$ accordingly increases.

\subsection{ZA versus True Solution}
\label{s33}

Equation (\ref{eq27}) is numerically solved to obtain the evolution of the homogeneous spheroid. For simplicity, we adopt the Einstein-de Sitter model ($\Omega = 1$ and $H = 2/3t$), but the results could qualitatively apply to any cosmological model. The initial axis ratio $\alpha _{1, {\rm in}} / \alpha _ {3, {\rm in}}$ or $\alpha _ {3, {\rm in}} / \alpha _ {1, {\rm in}}$ is 2, 4, or 8. The initial density contrast is $\delta_{\rm e, in} = D_{\rm in}$. The initial velocity $\dot{\alpha} _{i, \rm in}$ is determined using the ZA (eq. [\ref{eq28}]). We also obtain the evolution in the ZA.

Table 1 shows the scale factor at the moment of complete collapse $a_{\rm col}$. It is evident that, in a lower dimensional collapse, the ZA provides a more accurate result.

Figures 1 and 2 show the results for oblate collapses $(\alpha _ 1 = \alpha _ 2 > \alpha _3)$ and for prolate collapses $(\alpha _ 1 = \alpha _ 2 < \alpha _3)$, respectively, as a function of the normalized scale factor $a/a_{\rm col}$ ({\it solid lines}). For reference, the solutions for the one-dimensional planar, two-dimensional cylindrical, and three-dimensional spherical collapses are also shown ({\it dotted lines}).

First, we discuss the evolution of the half-lengths (Figs. 1$a$ and 2$a$). The collapse is rapid along the minor axis, while it is slow along the major axis. In the oblate cases (Fig. 1$a$), the collapse asymptotically becomes one-dimensional. The collapse along the minor axis is intermediate between the one- and three-dimensional collapses. If the initial axis ratio $\alpha _{1, {\rm in}} / \alpha _ {3, {\rm in}}$ is large, the collapse is close to the one-dimensional collapse. In the prolate cases (Fig. 2$a$), the collapse asymptotically becomes two-dimensional. The collapse along the minor axis is intermediate between the two- and three-dimensional collapses. If the initial axis ratio $\alpha _ {3, {\rm in}} / \alpha _ {1, {\rm in}}$ is large, the collapse is close to the two-dimensional collapse.

Second, for the gravitational acceleration along the minor axis, we compare the ZA with the true numerical solution (Figs. 1$b$ and 2$b$). In the oblate cases (Fig. 1$b$), the ZA's accuracy is intermediate between those in the one- and three-dimensional collapses. If the initial axis ratio is large, the ZA's accuracy is close to that in the one-dimensional collapse in which the ZA is identical to the true solution. This is because, as shown in Figure 1$a$, the collapse is close to the one-dimensional collapse. In the prolate cases (Fig. 2$b$), the ZA's accuracy is intermediate between those in the two- and three-dimensional collapses. If the initial axis ratio is large, the ZA's accuracy is close to that in the two-dimensional collapse. This is because, as shown in Figure 2$a$, the collapse is close to the two-dimensional collapse. The relative accuracy is the same if the ZA is compared with the true numerical solution along the density contrast $\delta_e$ in Figure 3.

Third, for the density contrast, we compare the ZA with the true numerical solution (Figs. 1$c$ and 2$c$). In the oblate cases (Fig. 1$c$), the ZA's accuracy is intermediate between those in the one- and three-dimensional collapses. If the initial axis ratio is large, the ZA's accuracy is close to that in the one-dimensional collapse. In the prolate cases (Fig. 2$c$), the ZA's accuracy is intermediate between those in the two- and three-dimensional collapses. If the initial axis ratio is large, the ZA's accuracy is close to that in the two-dimensional collapse. The ZA is less accurate in the density contrast than in the gravitational acceleration at the late stage of the evolution. This is because complete collapse occurs earlier in the true numerical solution than in the ZA (Table 1).

\section{DISCUSSION}
\label{s4}

The ZA is the linear approximation to replace $- {\bf \nabla}_x \Phi$ with $3 H^2 \Omega {\bf \Psi}/2$ for the gravitational acceleration in the Lagrangian equation of motion (eqs. [\ref{eq2}] and [\ref{eq6}]). This approximation becomes progressively more accurate in the order of three-, two-, and one-dimensional collapses (eq. [\ref{eq18}]). The reason is that, in a lower dimensional collapse, the gravity $1/\vert \mbox{\boldmath $x$} \vert ^{n-1}$ is more independent of the distance $\vert \mbox{\boldmath $x$} \vert$, and the change of the integrated density contrast $(\vert \mbox{\boldmath $x$} \vert ^n - \vert \mbox{\boldmath $q$} \vert ^n)/n$ is closer to the displacement $\Psi = \vert \mbox{\boldmath $x$} \vert - \vert \mbox{\boldmath $q$} \vert$. For the one-dimensional collapse, the ZA's gravitational acceleration is exact.

The spheroidal collapse preferentially proceeds along the minor axis, i.e., the direction of the strongest gravitational acceleration (Figs. 1 and 2). While an oblate spheroid asymptotically undergoes a one-dimensional collapse, a prolate spheroid asymptotically undergoes a two-dimensional collapse. The dimension of the collapse decreases in the course of the collapse. This is well reproduced in the ZA. The ZA's gravitational acceleration remains close to the true gravitational acceleration.

During structure formation in the real universe, an overdense region preferentially collapses along the direction of the strongest gravitational acceleration. The collapse asymptotically becomes one-dimensional as in the case of oblate spheroidal collapse.\footnote{
Since, at a later stage, the overdense region could collapse along the other directions, the final structure is not necessarily plane-like but could be point- or line-like (Arnold et al. 1982; see also Sahni \& Coles 1995; Kurokawa et al. 2001). The ZA is not accurate, because of the shell crossing, after the complete collapse along the direction of the strongest gravitational acceleration.}

Therefore, the ZA is accurate for the following reasons. First, the ZA is more accurate in a lower dimensional collapse (see also Yoshisato et al. 1998). Second, any gravitational collapse preferentially proceeds along the direction of the strongest gravitational acceleration, and it becomes progressively lower dimensional with the passage of time. That is, the ZA is accurate because the essence of the gravitational collapse is incorporated in the ZA.

This conclusion applies to other approximations based on the ZA, e.g., the truncated ZA (Coles et al. 1993). Such an approximation has inherited the ZA's dependence of accuracy on the dimensionality of the gravitational collapse (see also Yoshisato et al. 1998).

Our study is based on Buchert (1989), in which the ZA's gravitational acceleration was found to be exact for a one-dimensional collapse. This finding was for local motions. We have shown that this is also true for global motions. We have also shown that the ZA's gravitational acceleration is closer to the true gravitational acceleration in a lower dimensional collapse.

The usual explanation for the ZA's accuracy, especially in the density, is that the density contrast is a nonlinear function of the displacement (Catelan 1995; Bernardeau et al. 2002). Even if the displacement is small and lies in the linear regime, the corresponding density contrast could be large and lie in the nonlinear regime.  For a given change of the displacement, the corresponding change of the density contrast is larger in a higher dimensional collapse (eq. [\ref{eq18}]). The ZA should be more accurate in a higher dimensional collapse, according to this usual explanation. However, in practice (Figs. 1--3; see also Yoshisato et al. 1998), the ZA is more accurate in a lower dimensional collapse.

The ZA has long been known to be exact for a one-dimensional collapse (Doroshkevich et al. 1973). However, with this property alone, it is impossible to explain the ZA's accuracy. This is because, in general, gravitational collapse is not exactly one-dimensional. Our study has clarified that, even if the gravitational collapse is not exactly one-dimensional, the ZA is sufficiently accurate. Thus we have successfully explained why the ZA is accurate.

\acknowledgments

The authors are grateful to the referee for careful reading of the manuscript and helpful comments.

\begin{deluxetable}{lll}

\tablenum{1}
\tablecolumns{3}
\tablewidth{0pc}
\tablecaption{Scale Factor $a_{\rm col}$ at the Moment of Complete Collapse}

\tablehead{
\colhead{Initial Shape} &
\colhead{True}          &
\colhead{ZA} }

\startdata
three-dimensional spherical                    & 1.686 & 3 \\
two-dimensional cylindrical                    & 1.466 & 2 \\
one-dimensional planar                         & 1     & 1 \\
\hline
oblate        \phn $\alpha _1 / \alpha _3 = 2$   & 1.412 & 1.897 \\
\phm{oblate}  \phn $\alpha _1 / \alpha _3 = 4$   & 1.216 & 1.421 \\
\phm{oblate}  \phn $\alpha _1 / \alpha _3 = 8$   & 1.110 & 1.204 \\
prolate       \phn $\alpha _3 / \alpha _1 = 2$   & 1.577 & 2.420 \\
\phm{prolate} \phn $\alpha _3 / \alpha _1 = 4$   & 1.510 & 2.163 \\
\phm{prolate} \phn $\alpha _3 / \alpha _1 = 8$   & 1.480 & 2.059
\enddata
\end{deluxetable}

\notetoeditor{This table is to be placed in a single narrow column.}

\clearpage

\begin{figure}
\begin{center}
\resizebox{11cm}{!}{\includegraphics*[3cm,4cm][18cm,28cm]{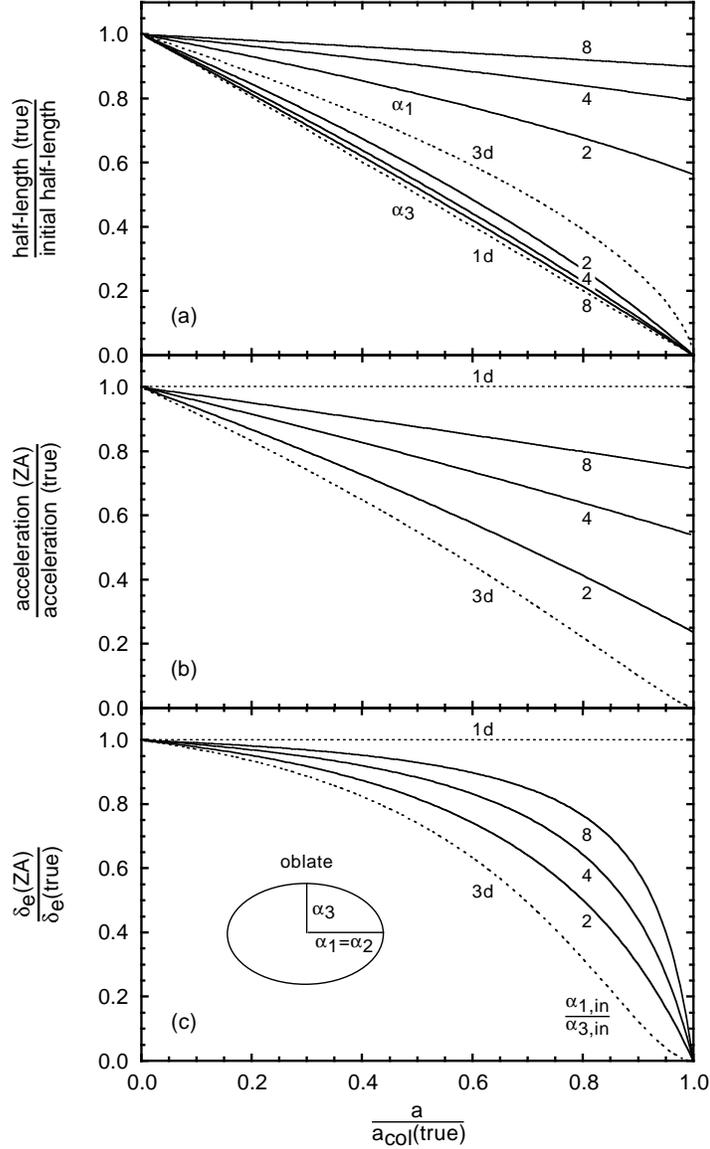}}

\caption[Fig1]{Oblate spheroidal collapse. ($a$) Half-lengths $\alpha _1$ and $\alpha _3$ in the true solution. ($b$) The ZA's gravitational acceleration along the minor axis $3 H^2 \Omega (\alpha _3 - \alpha _{3,{\rm in}})/2$ in the ZA. ($c$) Density contrast $\delta _{\rm e}$ in the ZA. The half-lengths are normalized by their initial values. The ZA's gravitational acceleration and density contrast in the ZA are respectively normalized by the true gravitational acceleration $-3 H^2 \Omega \delta _e A_3 \alpha _3 /4$ and density contrast in the true solution. The abscissa is the scale factor $a$ normalized by its value at the moment of complete collapse $a_{\rm col}$ in the true solution. The dotted lines denote the one- and three-dimensional collapses.}

\end{center}
\end{figure}

\clearpage

\begin{figure}
\begin{center}

\resizebox{11cm}{!}{\includegraphics*[3cm,4cm][18cm,28cm]{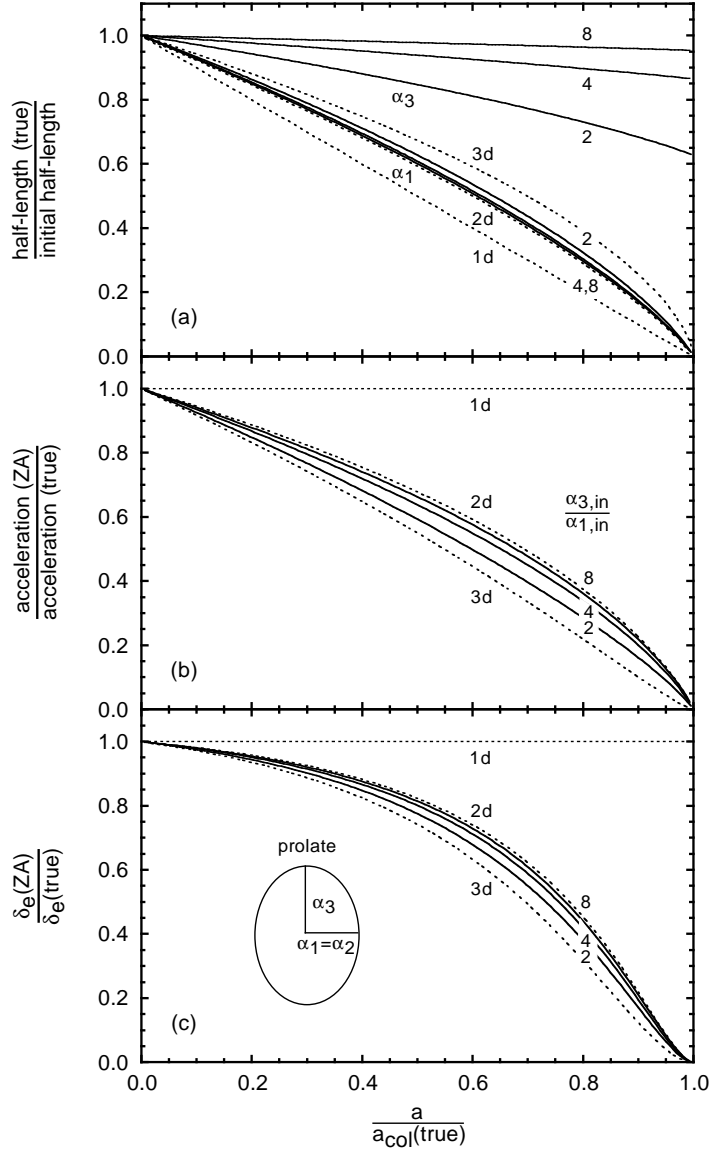}}

\caption[Fig2]{Same as Fig. 1 but for prolate spheroidal collapse. The ZA's gravitational acceleration is $3 H^2 \Omega (\alpha _1 - \alpha _{1,{\rm in}})/2$ and the true gravitational acceleration is $-3 H^2 \Omega \delta _e A_1 \alpha _1 /4$ along the minor axis. The dotted lines denote the one-, two-, and three-dimensional collapses.}

\end{center}
\end{figure}

\begin{figure}
\begin{center}

\resizebox{11cm}{!}{\includegraphics*[3cm,10cm][18cm,28cm]{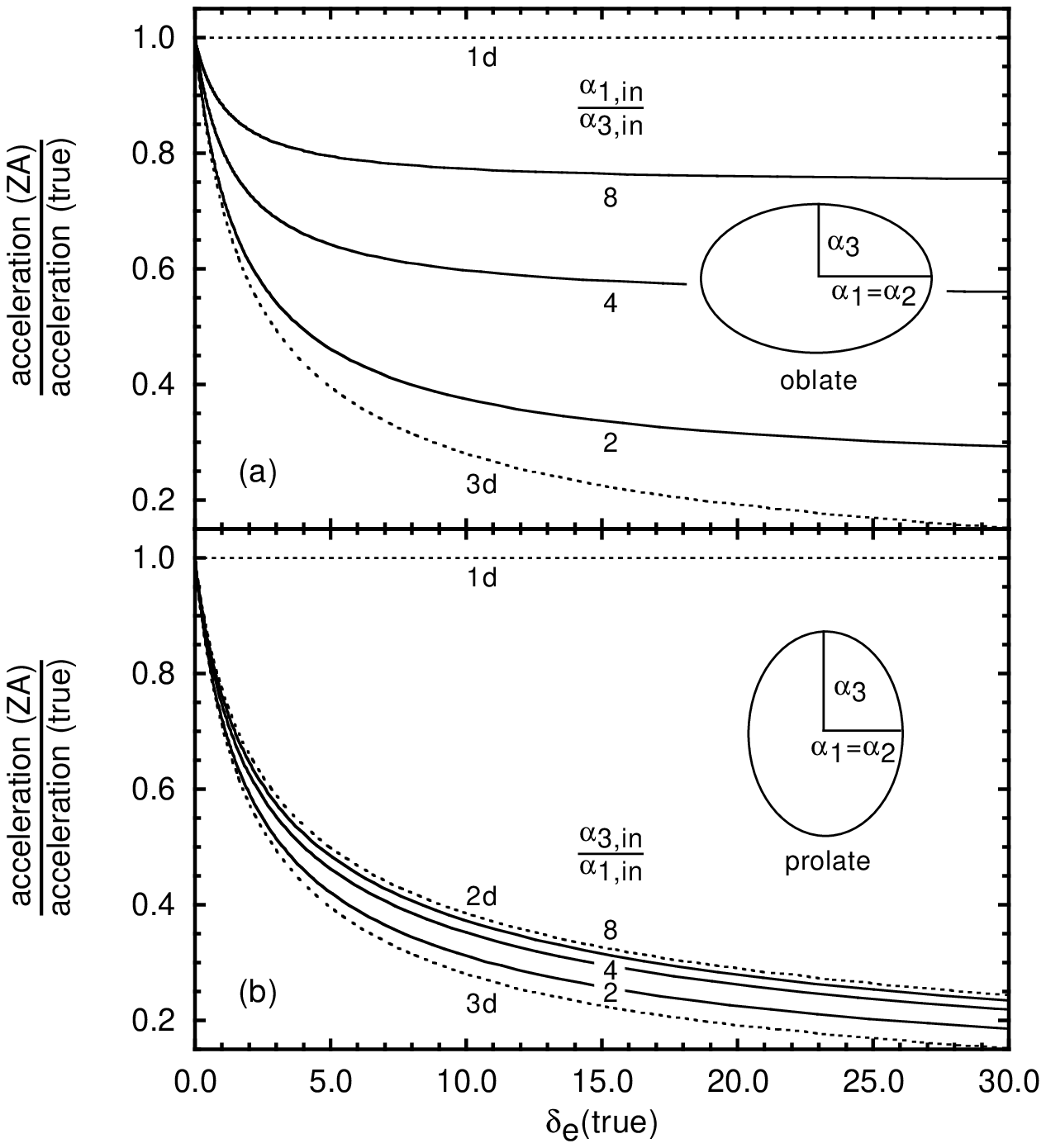}}

\caption[Fig3]{The ZA's gravitational acceleration along the minor axis normalized by the true gravitational acceleration. ($a$) Oblate spheroidal collapse. ($b$) Prolate spheroidal collapse. These are the same as Fig. 1$b$ and 2$b$, respectively, but the abscissa is the density contrast $\delta _{\rm e}$ in the true solution. }

\end{center}
\end{figure}


\begin{references}
\reference{} Arnold, V. I., Shandarin, S. F., \& Zel'dovich, Ya. B. 1982, Geophys. Astrophys. Fluid Dyn., 20, 111

\reference{} Bernardeau, F. 1994, ApJ, 427, 51

\reference{} Bernardeau, F., Colombi, S., Gazta\~naga, E., Scoccimarro, R. 2002, Phys. Rep., 367, 1

\reference{} Bernardeau, F., Singh, T. P., Banerjee, B., \& Chitre, S. M. 1994, MNRAS, 269, 947

\reference{} Bertschinger, E. 1998, ARA\&A, 36, 599

\reference{} Binney, J. 1977, ApJ, 215, 492

\reference{} Binney, J., \& Tremaine, S. 1987, Galactic Dynamics (Princeton: Princeton Univ. Press)

\reference{} Bouchet, F. R., Colombi, S., Hivon, E., \& Juszkiewicz, R. 1995, A\&A, 296, 575

\reference{} Buchert, T. 1989, A\&A, 223, 9

\reference{} Buchert, T. 1994, MNRAS, 267, 811

\reference{} Catelan, P. 1995, MNRAS, 276, 115

\reference{} Coles, P., Melott, A. L., \& Shandarin, S. F. 1993, MNRAS, 260, 765

\reference{} Dekel, A. 1994, ARA\&A, 32, 371

\reference{} Doroshkevich, A. G., Kotok, E. V., Novikov, I. D., Polyudov, A. N., Shandarin, S. F., \& Sigov, Yu. S. 1980, MNRAS, 192, 321

\reference{} Doroshkevich, A. G., Ryabenkii, V. S., \& Shandarin, S. F. 1973, Astrofizika, 9, 257

\reference{} Gurbatov, S. N., Saichev, A. I., \& Shandarin, S. F. 1989, \mnras, 236, 385

\reference{} Kofman, L., Bertschinger, E., Gelb, J. M., Nusser, A., \& Dekel, A. 1994, ApJ, 420, 44

\reference{} Kurokawa, T., Morikawa, M., \& Mouri, H. 2001, A\&A, 370, 358

\reference{} Lin, C. C., Mestel, L., \& Shu, F. H. 1965, ApJ, 142, 1431

\reference{} Matsubara, T., Yoshisato, A., \& Morikawa, M. 1998, ApJ, 504, 7

\reference{} Munshi, D., Sahni, V., \& Starobinsky, A. A. 1994, ApJ, 436, 517

\reference{} Munshi, D., \& Starobinsky, A. A. 1994, ApJ, 428, 433

\reference{} Peacock, J. A. 1999, Cosmological Physics (Cambridge: Cambridge Univ. Press)

\reference{} Sahni, V., \& Coles, P. 1995, Phys. Rep., 262, 1

\reference{} Sahni, V., \& Shandarin, S. 1996, MNRAS, 282, 641

\reference{} Shandarin, S. F., \& Zel'dovich, Ya. B. 1989, Rev. Mod. Phys., 61, 185

\reference{} Yoshisato, A., Matsubara, T., \& Morikawa, M. 1998, ApJ, 498, 48

\reference{} Yoshisato, A., Morikawa, M., \& Mouri, H. 2003, MNRAS, 343, 1038

\reference{} Zel'dovich, Ya. B. 1970, A\&A, 5, 84

\end{references}
\end{document}